\documentclass[aps,prl,reprint,showpacs]{revtex4-1}%
\usepackage{amssymb}
\usepackage{amsmath}
\usepackage{eurosym}
\usepackage{color}
\usepackage{graphicx}

\usepackage{dcolumn}
\usepackage{bm}
\usepackage{amsfonts}
\providecommand{\U}[1]{\protect\rule{.1in}{.1in}}
\begin{document}
\title{Nonequilibrium dynamics induced by scattering forces for optically trapped
nanoparticles in strongly inertial regimes}
\author{Yacine Amarouchene$^{1}$}
\email{yacine.amarouchene@u-bordeaux.fr }
\author{Matthieu Mangeat$^{1}$, Benjamin Vidal Montes$^{1}$, Lukas Ondic$^{2} $,
Thomas Gu\'{e}rin$^{1}$, David S. Dean$^{1}$}
\author{Yann Louyer$^{1}$}
\email{yann.louyer@u-bordeaux.fr }
\affiliation{(1) LOMA, CNRS UMR 5798, University of Bordeaux, F-33400 Talence, France}
\affiliation{(2) Institute of Physics, Academy of Sciences of the Czech Republic, CZ-162 00
Prague, Czech Republic.}
\date{\today}

\begin{abstract}
The forces acting on optically trapped particles are commonly assumed to be conservative.  
Nonconservative scattering forces induce toroidal currents in overdamped liquid environments, with negligible effects on position fluctuations. However, their impact in the underdamped regime remains unexplored. Here, we study the effect of nonconservative scattering forces on the underdamped nonlinear dynamics of trapped nanoparticles at various air pressures. These forces induce significant low-frequency position fluctuations along the optical axis and the emergence of toroidal currents in both position and velocity variables. Our experimental and theoretical results provide fundamental insights into the functioning of optical tweezers and a means for investigating nonequilibrium steady states induced by nonconservative forces.
\end{abstract}

\maketitle

Classical statistical mechanics establishes that a particle of mass $m$ subject to a potential $V$ in the presence of a heat bath, for instance the surrounding medium such as air or water will, in thermal equilibrium, be described by the Gibbs-Boltzmann probability distribution for its position ${\bf x}$ and velocity ${\bf v}$ \cite{mac}
\begin{equation}
P_{GB}({\bf x},{\bf v}) = \frac{1}{Z_{xv}}\exp\left(-\frac{\frac{1}{2}m{\bf v}^2 + V({\bf x})}{k_BT}\right).\label{gb}
\end{equation}
Here, $T$ is the temperature imposed by the heat bath, $k_B$ is Boltzmann's constant and $Z_{xv}$ a normalization constant, better known as the canonical partition function. Eq.~(\ref{gb}) has some remarkable consequences that are today taken for granted. First, the position ${\bf x}$ and the velocity ${\bf v}$ are independent random variables. The marginal distribution for the velocity is the celebrated Maxwell-Boltzmann distribution, proposed at the very inception of the field of statistical mechanics, and is independent of the interaction potential. Another remarkable property is that the equilibrium distribution is independent of the dynamics. This, today obvious, observation means that at the same temperature and potential, a particle trapped in water will have the same equilibrium distribution as the one trapped in air.  Another feature of the Gibbs-Boltzmann distribution is that it does not have any currents in position or velocity, this is necessary in an equilibrium distribution so that it satisfies time reversal symmetry \cite{ruelle}.

\begin{figure}[b]
\begin{center}
\includegraphics[width=85mm]{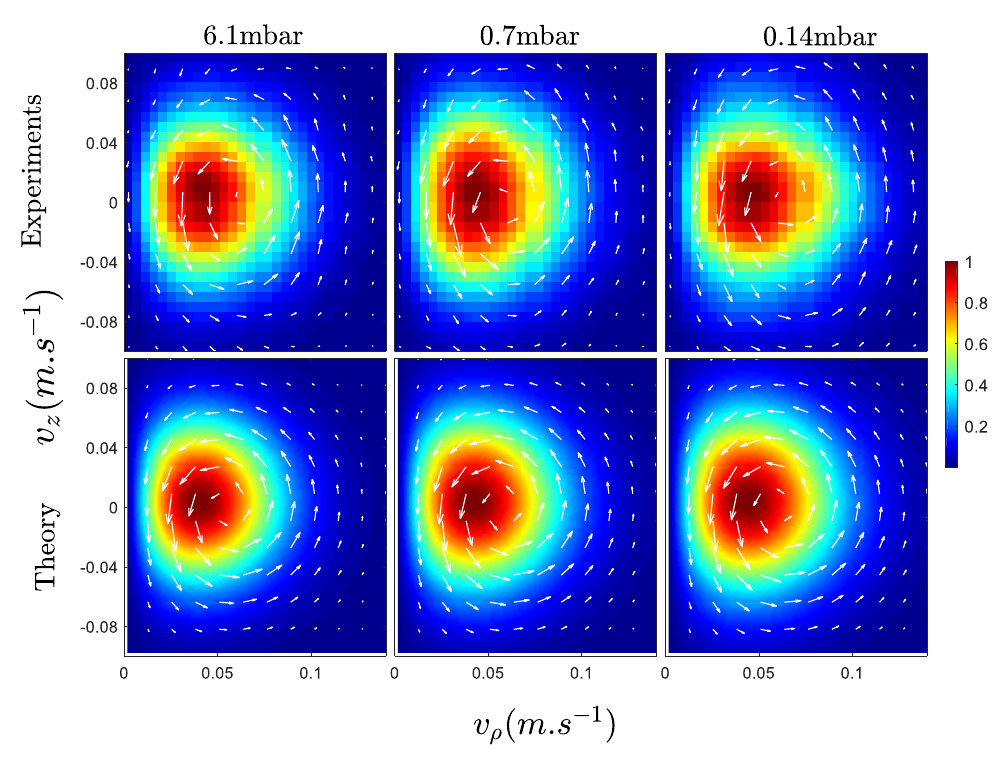}
\end{center}
\caption{Experimental and theoretical probability currents $\overline {\bf J}_{\bf v}$ in velocity space for several pressures in the (radial~$v_\rho$, axial~$v_z$) plane. The theory is described in the companion article \cite{PaperTheo}. The color bar displays the steady state probability distribution $P_s\left( {\bf v} \right) $.}
\label{Fig1}
\end{figure}

However, when the force acting on the particle is not derived from a potential, much less is known on the nonequilibrium stationary distribution. 
Optically trapped particles have recently attracted attention as a model system to study nonequilibrium forces \cite{Florin, Roichman, BoSun, GrierTheo,Laporta,Pesce}. As originally shown by Ashkin \cite{Ashkin}, a particle can be trapped by using the (conservative) intensity gradient force of a laser. It is also well known that an optically trapped particle is also submitted to \textit{nonconservative} scattering forces. These forces, first measured in Ref. \cite{Florin}, induce steady-state nonequilibrium probability currents, which develop over time due to thermal fluctuations displacing the particle away from its stable mechanical equilibrium point into the nonconservative force field. Such currents have been demonstrated % both experimentally and theoretically
only in the \textit{overdamped regime} \cite{Roichman, BoSun, GrierTheo} where the effect of nonconservative forces turns out to be modest \cite{Laporta,Pesce}. 
In the underdamped regime, trapping by intensity gradient forces has recently been used in vacuum \cite{Raizen,Novotny}, leading to impressive proposals and experimental results in ultra-weak force sensing and fundamental tests of quantum mechanics \cite{Arvanitaki,Lochan,Bateman,Moore,Ranjit,Rider,Geraci,Liu}. These studies consider only conservative forces, so that nonequilibrium statistical mechanics of inertial (underdamped) optically trapped particles remains unexplored. 

In this Letter, we describe an experimental and theoretical study of an underdamped particle in an optical trap which generates a well characterized nonconservative force. We show that probability currents exist in both position and velocity space, the latter being inaccessible in the overdamped case. These currents means that even the Maxwell-Boltzmann distribution is modified by nonconservative forces (see Fig.~\ref{Fig1}). Furthermore, these modifications are shown to depend on the details of the dynamics, in this case the damping. As well as modification of static quantities, we also show that the resulting steady state \textit{dynamics} is drastically modified by the nonconservative force. Namely, we observe, for the first time in an optical trap, pressure-dependent additional low-frequency broadband axial positional fluctuations. Such
effects dominate over thermal fluctuations and are well too weak to be observed in the overdamped regime \cite{Laporta, Pesce}. Such low-frequency contributions in the axial motion originate from the scattering forces and are further amplified by the nonlinear gradient components. This Letter is accompanied by a theoretical article \cite{PaperTheo}. 
 
We first describe our experiment (detailed in the Supplemental Material - SM \cite{SM}), which  consists of trapping, in a near vacuum, a $68$-nm radius fused silica nanoparticle with a $430$-mW tightly focused  linearly polarized laser beam at $\lambda=1064$~nm by using a high $0.8$ numerical aperture objective. The Gaussian laser beam propagates in the $z$ direction, while $x,y$ are the transverse directions. The pressure can be varied to modify the friction coefficient applied on the bead and thus to explore the effect of inertia on the particle's motion. A new calibration procedure that takes into account the nonlinear aspects of the optical trap is also presented in SM \cite{SM}. This enables us to estimate the geometrical parameters of the trap, the intrinsic damping rate $\Gamma$, but also the size of the trapped particle and center of mass motion temperature $T$. The final trap parameters obtained from this procedure are: $w_{x}=0.915$~$\mu$m, $w_{y}=1.034$~$\mu$m and $w_{z}=2.966$ $\mu$m (beam waist radii $w_{i=x,y}$ and Rayleigh length $w_z$). In the following, these parameters are used to compare the theoretical and experimental results. 

A crucial advantage of the underdamped system  studied here  is that experimental resolution allows an accurate and unambiguous determination of both the velocity ${\bf V}_t$ and the acceleration $\dot {\bf V}_t$, as compared to overdamped systems \cite{Roichman, BoSun, GrierTheo}. This enables us to measure the currents both in position and velocity space. By definition, the probability density function $P({\bf x},{\bf v},t)$ is simply given by the average over stochastic trajectories $P({\bf x},{\bf v},t)=\langle \delta({\bf x}-{\bf X}_t)\delta({\bf v}-{\bf V}_t)\rangle$, with ${\bf X}_t,{\bf V}_t$ the instantaneous position and velocity of the particle at time $t$, and $\langle \cdot \rangle$ denoting ensemble averaging. The currents ${\bf J}_{\bf x}$ and ${\bf J}_{\bf v}$ in position and velocity space are defined by considering the general transport equation for this system given by $\partial_t  P({\bf x},{\bf v},t)= -\nabla_{\bf x} \cdot{\bf J}_{\bf x} -\nabla_{\bf v} \cdot{\bf J}_{\bf v}$. 
Integrating these currents leads to the definition of effective currents $\overline {\bf J}_{\bf x}= \int d{\bf v} {\bf J}_{\bf x} = \langle  {\bf V}_t \delta({\bf x}-{\bf X}_t)\rangle$ in position space and $\overline {\bf J}_{\bf v}= \int d{\bf x} {\bf J}_{\bf v} = \langle \dot {\bf V}_t \delta({\bf v}-{\bf V}_t) \rangle$ in velocity space. These currents are experimentally estimated either by a six-dimensional
histogram binning \cite{Florin} or kernel density techniques \cite{BoSun}. Here, steady state is reached by using suitably small bin sizes and time averaging.

We consider the marginal probability distribution of the velocity and its associated current, which to our knowledge is measured here for the first time. Shown in Fig.~\ref{Fig1} is the 
steady current in velocity space  $\overline {\bf J}_{\bf v}$ measured at different pressures (and therefore different friction coefficients). We see that the current (shown as arrows) is non-zero - a clear indication of the presence of a nonconservative force and deviation from the Maxwell Boltzmann distribution. Interestingly, these currents seem to form rolls (in velocity space) that are in the opposite direction to currents in position space (Fig.~2, inset). 

To get a theoretical understanding of these currents, we consider the Langevin dynamics of  a trapped particle,  
\begin{align}
m \ddot{x_i}+m \Gamma \dot{x_i}   = -\kappa_i x_i+ F_i^\mathrm{Duffing}+F_i^\mathrm{scatt}+f_i(t) \label{EqLangevin}
\end{align}
with $m$ the particle mass, $m \Gamma$ the friction coefficient, $f_i(t)$ is the thermal white noise force \cite{PaperTheo}. In addition to the standard harmonic force $-\kappa_i x_i$ (with $\kappa_i$ the stiffness in the direction $i$), the trap exerts two additional forces on the particle, which we calculate at next-to-leading order for small displacements relative to the laser wavelength \cite{GieselerNature}.  
The term $F^\mathrm{Duffing}$ contains the cubic nonlinearities arising from the gradient force and is widely studied in the one-dimensional Duffing oscillator \cite{DykmanNL,Renz1}:
\begin{align}
F_i^\mathrm{Duffing}= \kappa_i x_i\left[(1+\delta_{i,z})\left( \frac{2x^2}{w_x^2}+\frac{2y^2}{w_y^2}\right)+ \frac{2z^2}{w_z^2}    \right]. 
\end{align}
The  main contribution of the scattering force is in the axial direction and is given by
\begin{equation}
F_{i}^{\text{scatt}}=\delta_{iz}  \frac{\alpha^{\prime\prime}}{\alpha^{\prime}}\kappa_{z}\left(  \gamma_{0}+\sum_{j=x,y,z}\gamma_{j}x_{j}^{2}\right),
	\label{FscattFull}
\end{equation} 
where $\alpha^{\prime}$ and $\alpha^{\prime\prime}$ are, respectively, the real and imaginary parts of the effective polarizability $\alpha=\alpha_{0}/\left[  1-ik^{3}\alpha_{0}/(6\pi\epsilon_{0})\right]$, with $k$ the laser wavevector and $\epsilon_{0}$ the vacuum permittivity \cite{GieselerNature}. Optical forces arise from the interaction between the electromagnetic field of the Gaussian beam and the polarizability $\alpha_{0}=4\pi\epsilon_{0}R_{p}^{3}\left( \epsilon-1\right)  /\left(  \epsilon+2\right)  $ of a spherical object of dielectric constant $\epsilon$ and radius $R_p$. The coefficients $\gamma_i $ depend on the geometrical parameters of the focused laser:~$\gamma_{0}=w_{z}\left( w_{z}k-1\right)  $, $\gamma_{i=x,y}=k/2-2\gamma_0/w_{i}^{2}$ and $\gamma_{z}=\left(  2-w_{z}k\right)  /w_{z}$ \cite{GieselerNature}. 
 
Importantly, these scattering forces are nonconservative. In the overdamped regime, these forces are known to give rise to nonequilibrium probability currents that have been termed Brownian vortices, and arise in a  \textit{minimal scattering model} (MSM), which assumes an axisymmetric trap  ( \textit{i.e.}, $\gamma_x=\gamma_y$ and $\kappa_x=\kappa_y=\kappa_\perp$), disregards  Duffing nonlinearities  as well as the axial term ($\gamma_z z^2$) in the scattering force. In the MSM, this force can then be written as
\begin{align}
F_{i}^{\text{scatt}}=\delta_{iz}  \varepsilon \kappa_\perp a \left[1- (x^2+y^2)/a^2\right],
\end{align}
where the dimensionless parameter $\varepsilon$, which quantifies the magnitude of scattering forces, and the length-scale $a\simeq w_\perp/\sqrt{2}$ at which scattering forces vary, are easily identified by comparing with (\ref{FscattFull}).

Until now, studies of this MSM model considered overdamped motion \cite{Laporta}, here we extend the  study to inertial particles. The theory to perform this task is described in the companion paper \cite{PaperTheo}. In this model the fluxes can be exactly computed at first order in $\varepsilon$, 
\begin{align}
&\overline{\bf J_x}=  \varepsilon A_x \left[\frac{\kappa_z}{\kappa_\perp} z \rho\hat{{\bf e}}_{\rho} +\left(\frac{2k_BT}{\kappa_\perp}-\rho^2\right)\hat{{\bf e}}_z\right] e^{-\frac{\kappa_\perp {\bf \rho}^2+\kappa_z z^2}{2k_BT}} \\
&\overline{\bf J_v}= -  \varepsilon A_v \left[ v_z {\bf v}_\rho +\left(\frac{2k_BT}{m}-{\bf v}_\rho^2\right)\hat{{\bf e}}_z\right]e^{-\frac{m {\bf v}^2}{2k_BT}}
\end{align}
where the analytical expressions for the (positive) amplitude factors $A_x$ and $A_v$ are explicitly given in \cite{PaperTheo}, with $\rho$ the distance to the optical axis and ${\bf v}_\rho$ the transverse component of the velocity. The geometry of fluxes in position space is the same as for overdamped systems, but interestingly the amplitude factor is found to be non-monotonic with the friction coefficient. A similar geometry, but with opposite sign, holds in velocity space. 

\begin{figure}[h]
\begin{center}
\includegraphics[width=85mm]{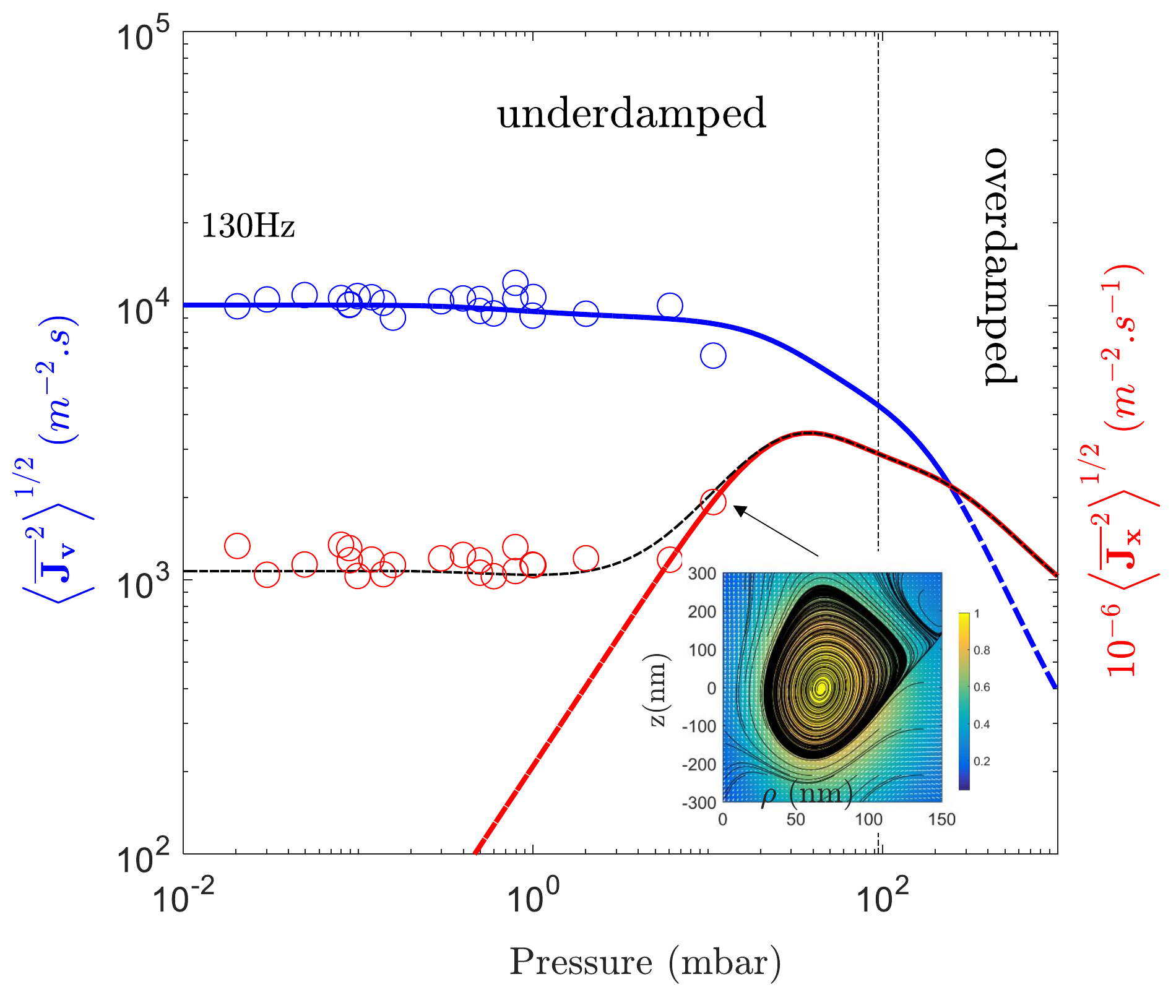}
\end{center}
\caption{In blue, experimental (circle) and theoretical \textit{minimal scattering model} MSM (line, \cite{PaperTheo}) probability current amplitudes $\left\langle \overline{\bf J_v}^2\right\rangle^{1/2}$ in velocity space for several pressures, obtained by a uniform average over the velocity window shown in Fig.~\ref{Fig1}.   
Similarly, experimental and theoretical MSM current amplitudes in the position space are shown in red. The theory uses parameters obtained from the calibration \cite{SM}. The dashed line is obtained from Eq.~(6) via an effective damping constant as seen in~\cite{SM}.
Also shown as an inset is the current map in position space measured at $10.8$ mbar.
}
\label{Fig2}
\end{figure}

In both position and velocity spaces, the predictions for the current geometry are very close to the experimental observations shown in Fig.~1 and Fig.~2 (inset). The comparison can be made quantitative by defining a scalar quantity characterizing the amplitude of the fluxes. We thus define $\left\langle \overline{\bf J_v}^2\right\rangle$ as the uniform average of the squared flux over the window of velocities shown in Fig.~1. We represent in Fig.~\ref{Fig2} the pressure dependence of the amplitude $\left\langle \overline{\bf J_v}^2\right\rangle^{1/2}$ and compare with the MSM  theory. As predicted by the theory \cite{PaperTheo}, the current's amplitude saturates at low pressure at a value that corresponds to the theory if one uses the parameter values arising from the calibration technique. We find $\varepsilon = 0.05$ consistent with the value of the imaginary index of fused silica. The corresponding circulation rate for these vortices in velocity space is about $\Omega_0=(1/2\pi)\int d{\bf v}({\bf \nabla}_{\bf v}\times {\bf J_v})\cdot {\bf e}_\theta\approx~130$~Hz. The quantitative success of the MSM theory, which neglects Duffing nonlinearities, is likely due to the fact that the marginal distribution is dominated by positions close to the trap center where nonlinear effects are by definition small. 
In position space, rather than  following the MSM theoretical prediction, we see that the amplitude $\left\langle \overline{\bf J_x}^2\right\rangle^{1/2}$ saturates at low pressure (Fig.~2). 
This shows that Duffing nonlinearities, absent in the MSM theory, are important to quantitatively describe fluxes in position space. Such nonlinearities lead to a saturation of effective damping rates seen in the spectral densities at the oscillator eigenfrequencies \cite{SM}. Interestingly, using this effective damping rate $\Gamma_{\texttt{eff}}$ (identified in Fig. S2(e) of \cite{SM}) leads to a good agreement with the data shown in Fig.~2. 
This, however, deserves further experimental and theoretical investigation.

In what follows, we present evidence that scattering forces make a dominant contribution to a dynamic quantity: the power spectral density (PSD) $S_{zz}(\omega)$ in the longitudinal direction. Within the MSM, $S_{zz}(\omega)$ displays a low-frequency peak and is exactly given for low pulsations $\omega$ by \cite{PaperTheo}~: 
\begin{equation}
S_{zz}\left(  \omega\right)  =\frac{2k_{B}T\Gamma}{m\Omega_{z}^{4}}+4\left(
\frac{\varepsilon k_{B}T}{am\Omega_{z}^{2}}\right)  ^{2}\frac{\Gamma}%
{\Gamma^{2}+\omega^{2}}, \label{SzzModel}%
\end{equation}
where $\Omega_z=(\kappa_z/m)^{1/2}$. The first term in the expression of $S_{zz}(\omega)$ corresponds to the thermal component and the second term is due to nonconservative forces. Its origin stems from the fact that  $x^2(t)$ and $y^2(t)$ contain weakly varying functions, giving rise to a Lorentzian low frequency component - this is similar to the analysis of Ref.~\cite{DykmanNL}. If we compare the two terms, the above equation implies that the low frequency part of the spectrum becomes increasingly dominated by the component due to scattering forces in the underdamped limit: this is a major difference with the overdamped case, where the effect of these scattering forces are insignificant at all frequencies \cite{Laporta}. 

However, Duffing nonlinearities are neglected in the MSM, which can appear as naive since these nonlinearities are well-known to be essential to describe one-dimensional underdamped oscillators \cite{GieselerNature}. To determine whether the above predictions hold in a more realistic model, we now turn to Langevin simulations of the 3D fully nonlinear Eq.~(\ref{EqLangevin}) and calculate the respective PSDs \cite{Sivak}. These simulations are performed for parameter values obtained by the trap calibration \cite{SM}.  Here, we distinguish between different situations:~ the \emph{Duffing} case, where we neglect the scattering force; the \emph{Fully nonlinear} model (FNL) includes all terms of Eq.~(\ref{EqLangevin}), and the analytically soluble MSM which neglects Duffing nonlinearities. 
We note that our numerical simulations reveal that the transverse scattering forces $F_{i=x,y}^{\text{scatt}}$ have a negligible contribution in the frequency and position fluctuations, which is why we omit them in Eq.~(\ref{FscattFull}).  

\begin{figure}[b]
\begin{center}
\includegraphics[width=85mm]{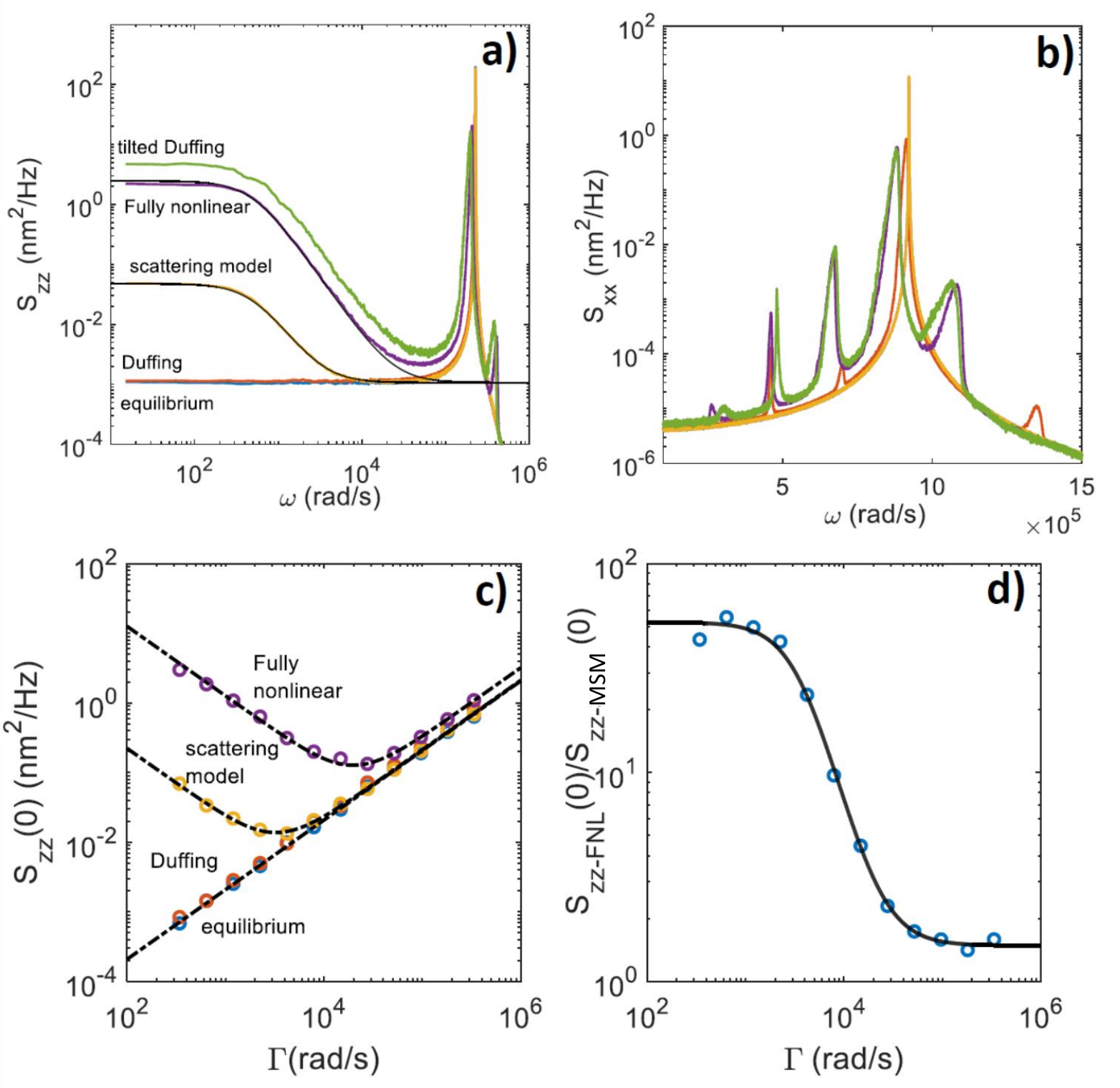}
\end{center}
\caption{Numerical simulation results using the waists $w_{i}$ extracted \ from
the nonlinear calibrations at a pressure of $0.1$mbar (\textit{i.e.} for $\Gamma=500$~rad$/$s). PSDs $S_{ii}$ along $i=z$ axis (a) and $i=x$ axis (b) for the different cases described in
the text (the same color legend shown in Fig. 1(a) is used for all the figures). c) $S_{zz}(0)$ versus damping rate $\Gamma$. d) Ratio of the
zero frequency response $S_{zz}(0)$ for the fully nonlinear model and
the MSM (see text). }
\label{Fig3}
\end{figure}

\begin{figure}[h]
\begin{center}
\includegraphics[width=85mm]{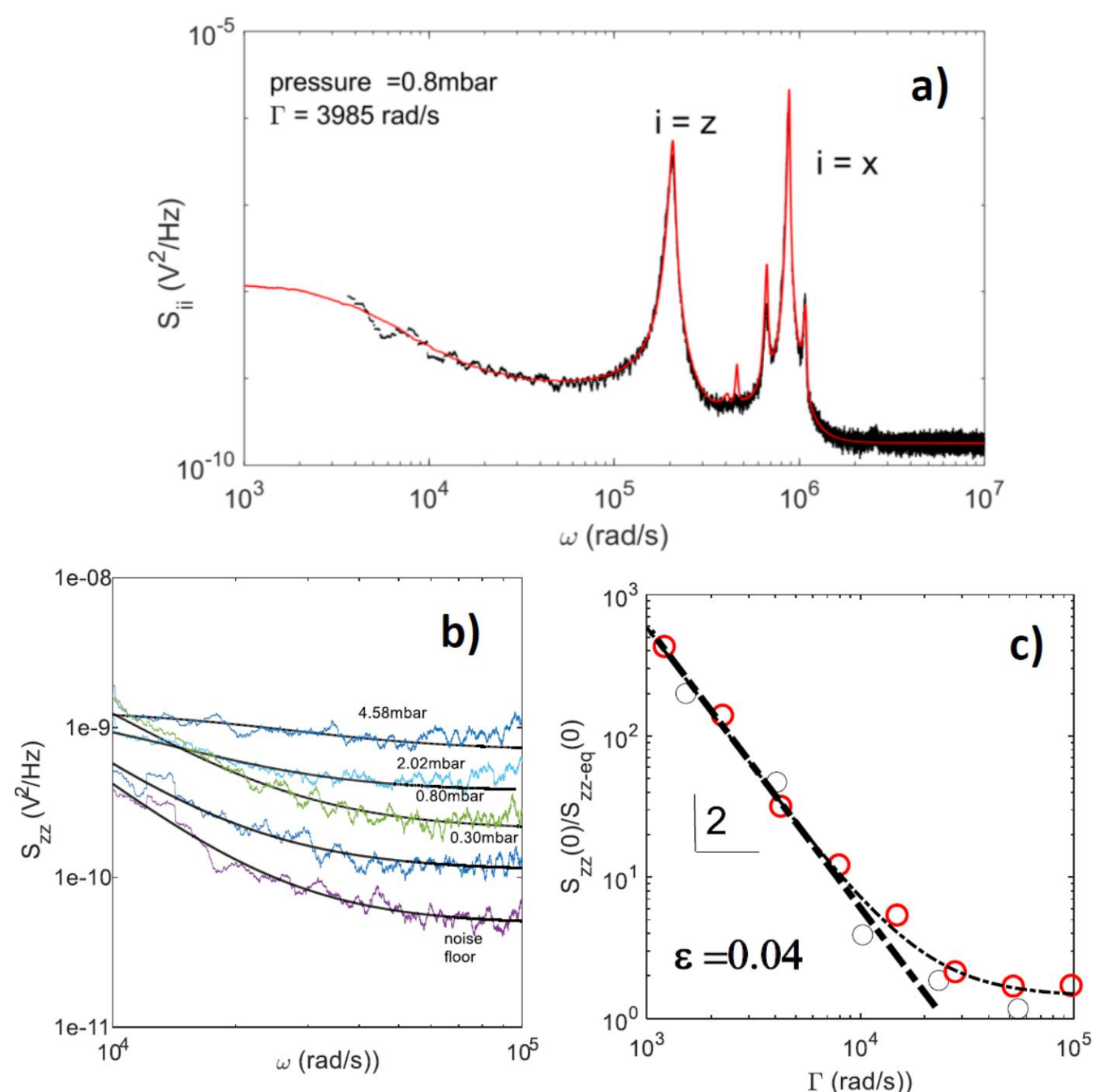}
\end{center}
\caption{a) Experimental PSDs \ (in directions $x$ and $z$) versus fully nonlinear simulation
of the PDSs at a pressure of 0.8 mbar [black experiments/red simulations, idem for (c)]. b) Low frequency part of $S_{zz}(\omega)$ at various pressures. The fits to the data correspond to a fit using the functional form of Eq. (\ref{SzzModel}). The deduced values of $S_{zz}(0)$ relative to the equilibrium value versus the damping rate $\Gamma$ is shown in c). The thin line is a guide to the eye
while the thick line highlights the $\Gamma^{2}$ scaling discussed in the text.}
\label{Fig4}                         
\end{figure}

Fig.~\ref{Fig3}(a) and \ref{Fig3}(b) show PSDs in the axial and transverse directions. We see how the scattering force increases the frequency shift of the resonance by a few percent.  
More importantly, the low-frequency peak is present in the MSM (as expected) but completely absent in the Duffing model, which shows that nonconservative forces are necessary for its emergence. In the complete nonlinear model, this peak is not only still present, but is also largely amplified (when compared to the MSM). 
Note that these low-frequency overdamped fluctuations are absent in the transverse spectral densities (data not shown). For comparison with previous works in analog circuits \cite{DykmanM0}, we also represent a \textit{tilted Duffing case}, in which case Duffing terms and the linear part of the scattering force in Eq.~(4) are used. This \emph{tilted Duffing}
model clearly overestimates the low-frequency response [Fig.~\ref{Fig3}(a)].
Eq.~(\ref{SzzModel}) perfectly reproduces the numerical simulation of $S_{zz}(\omega)$ for the MSM (see Fig.~\ref{Fig3}(a) and \cite{PaperTheo}) and shows that the low-frequency overdamped
component (only visible in the axial direction) has a corner frequency given
by $\Gamma$. The $\Gamma$ dependence of the amplitude $S_{zz}(0)$ also holds for the
different cases described above, as seen in Fig.~\ref{Fig3}(c). Strikingly, the theory
also gives good fitting results for the fully nonlinear model at low
frequency provided that a pressure dependent 
correction factor $S_{zz-\text{FNL}}(0)/S_{zz-\text{MSM}}(0)$ 
[shown in Fig.~\ref{Fig3}(d)] is used solely for the scattering
 term in Eq.~(\ref{SzzModel}). 

Finally, we investigate experimentally the presence of a low frequency component in the axial PSD due to scattering forces. In practice, the low-frequency $1/f$-like noise of the laser hinders the
direct observation of $S_{zz}(0)$. To circumvent this technical issue, Fig.~\ref{Fig4}(a) displays the raw experimental PSDs $S_{ii}\left(  \omega\right)
=c_{x}S_{xx}\left(  \omega\right)  +c_{z}S_{zz}\left(  \omega\right)  $ for a
limited range of frequencies at $0.8$~mbar ($c_{x}$ and $c_{z}$ being calibration
constants). These experimental data are in good agreement with our numerical
simulations using the calibration factors and the beam parameters \cite{SM}. Fig.~\ref{Fig4}(b) displays the pressure dependence of the low-frequency response of the PSDs along the $z$ axis. Knowing $\Gamma$ for each pressure,
fitting of Eq.~(\ref{SzzModel}) enables us to infer $S_{zz}\left(  0\right)  $
relative to its equilibrium value $S_{zz-eq}\left(  0\right)  $ as shown in
Fig.~\ref{Fig4}(c). The scaling $S_{zz}(0)/S_{zz-eq}(0)\propto \Gamma^{-2}$, which is expected from the MSM, is also observed at low pressures. 
Using corrections in temperature shown in Fig.~S2(d) of \cite{SM} and the correction factor of Fig.~\ref{Fig3}(d) provides a means to estimate $\varepsilon\approx0.04$ corresponding to an
imaginary refractive index of $\approx10^{-7}$, as expected for fused silica
\cite{Kitamura} and in agreement with the previous value of $\varepsilon$ used to describe the fluxes amplitude.

In summary, we have experimentally and theoretically demonstrated the effect
of radiation pressure for optically trapped nanoparticles in the nonlinear
underdamped regime. In a near vacuum environment, position fluctuations are
amplified at low pressure. Toroidal Brownian vortices in both position and velocity space have been observed. 
The currents in position space, however, deserve further theoretical and experimental study. In particular, understanding the
exact topology of Brownian vortices (in position-velocity space) and their
efficiencies versus dissipation require further experimental investigation in
the underdamped regime \cite{GrierTheo}, but also paves the way
towards studying the time reversal symmetry breaking induced by non
conservative forces \cite{PaperTheo}. Our work opens a new pathway for studying
nonequilibrium statistical physics for a wide range of damping regimes. It
also highlights the importance of fully characterizing optical traps in the underdamped regime that is relevant for the quest of ultra-weak force sensing and fundamental tests of quantum mechanics.

\begin{acknowledgments}
We thank L. Haelman for the construction of the vacuum chamber, W. Benharbone,
S. Cassagnere and B. Tregon from the Electronics Lab, and R. Avriller for
stimulating discussions. This work was partially funded by the Bordeaux IdEX
program - LAPHIA (ANR-10-IDEX-03-02) -Arts et Science (Sonotact 2017-2018) and
Projet Region Aquitaine (2018-1R50304).
\end{acknowledgments}

\pagebreak

\onecolumngrid
\begin{center}
\textbf{\large Supplemental Material:\\Nonequilibrium dynamics induced by scattering forces for optically trapped
nanoparticles in strongly inertial regimes}
\end{center}

\setcounter{equation}{0}
\setcounter{figure}{0}
\setcounter{table}{0}
\setcounter{page}{1}
\renewcommand{\theequation}{S\arabic{equation}}
\renewcommand{\thefigure}{S\arabic{figure}}
\renewcommand{\bibnumfmt}[1]{[S#1]}
\renewcommand{\citenumfont}[1]{S#1}

\setcounter{equation}{0}

\setcounter{figure}{0}
\makeatletter
\renewcommand{\theequation}{S\arabic{equation}}
\renewcommand{\thefigure}{S\@arabic\c@figure}
\renewcommand{\bibnumfmt}[1]{[S#1]}
\makeatother

Here we describe the experimental setup for the particle trap. We give details about the particle loading in the trap. We present a new nonlinear calibration method to extract trap and particle parameters from time traces and the fitting to the theoretical model to the experimental power spectral density (PSD).

\begin{figure}[h]
\begin{center}
\includegraphics[width=135mm]{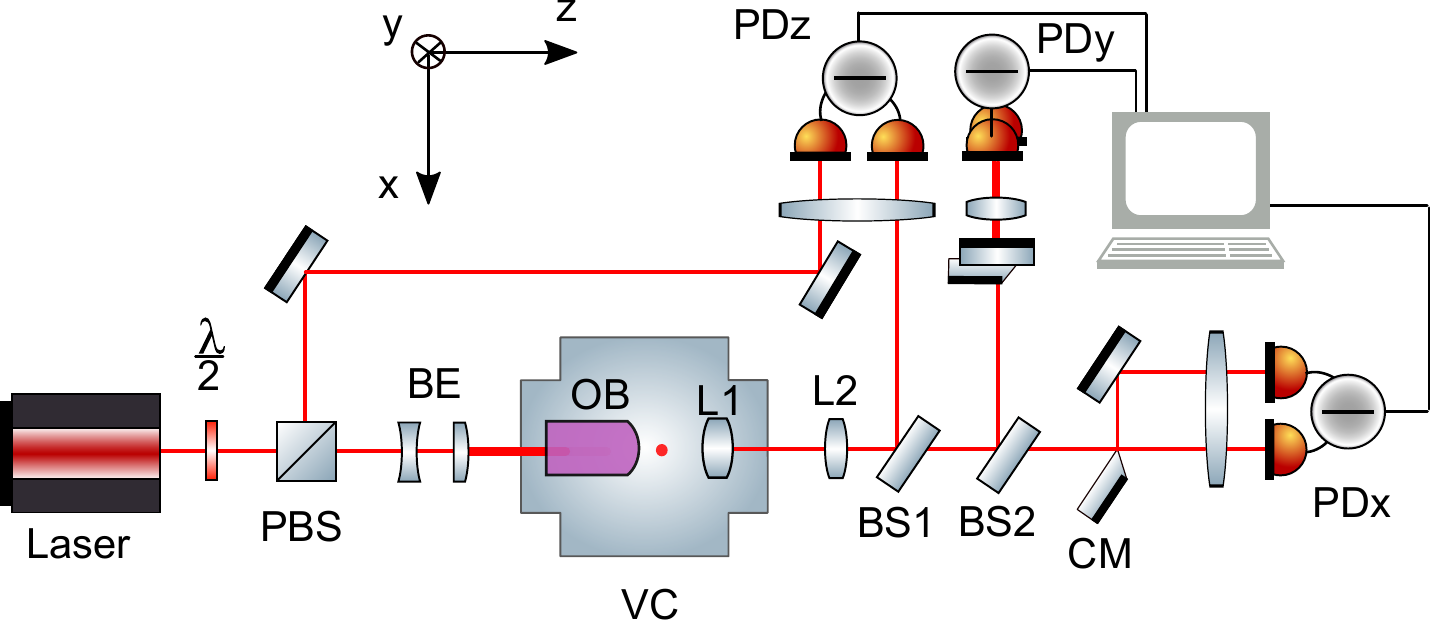}
\end{center}
\caption{Experimental setup (see text for details).}
\label{FigS1}
\end{figure}

\section{Experimental setup}

We construct, inside a vacuum chamber, a linearly polarized single-beam
gradient trap by focusing a low-noise $\lambda=1064$~nm laser (430 mW power)
with a high $0.8$ numerical aperture objective. Nebulization of a dilute
solution of fused silica nanoparticles (density, $\rho=2200$~$kg.m^{-3}$,
refractive index $n=1.45,$ with a nominal diameter of $136$ nm\ and a standard
deviation of $5$ nm) in ethanol leads to the trapping of a single particle at
ambient pressure. The particle's trajectory in $3D$ is measured by forward
scattered light interferometry [4] and recorded with a sample rate
of $10$~MS$/$s ($16$~bits of vertical resolution) with more than
$3\times10^{6}$ points for each axis.

Figure \ref{FigS1} displays all the optical elements used in this work (not indicated: a few density plates). The laser (Azurlight systems) embedding a Faraday Isolator has a relative intensity noise as low as $-220$ dBc$/$Hz at the oscillator eigenfrequencies and a polarization stability $<0.1~\%$. All the optical paths are minimized in order to avoid as much as possible air currents and effects of laser pointing fluctuations ($<\pm 0.5$ $\mu$rad over several hours). 

The half wave-plate ($\lambda/2$) and the polarizing beam splitter (PBS) regulate the optical trapping power between the two optical beams. BE is the beam expander, which slightly under fills the back aperture of the microscope objective (Nikon IR Plan-APO N.A.: $0.8$, w.d.: $1$ mm) used to trap particles. Under filling the back aperture reduces geometrical aberrations. Thus if we use a circularly polarized trapping beam, we can obtain an isotropic trap $w_x=w_y$ (not shown). L$1$ is an aspherical lens (numerical aperture N.A. of $0.8$, Edmund Optics) used to collect both the incident field and scattering light. L$2$ is a relay lens to image the back focal plane into the photodetectors. BS$1$ and BS$2$ are beam splitters for equally distributing the optical intensity to the InGaAs photodectors (Thorlabs PDB$425$C-AC). CM is a D-shaped mirror used to cut the beam in two halves part. A single lens is used to focus the two separated beams into the balanced photodiodes. PDx, PDy and PDz are used for detecting the three dimensional motion of the trapped particle. The exiting signals are then sent to a computer via a 16-bit DAQ board (Razor 16, Gage) recorded in binary files and analyzed using custom made routines.  

Silica nanoparticles (Microparticles GmbH) are loaded into the optical tweezer at atmospheric pressure with a nebulizer (Omron Micro-Air). The nebulizer is filled with a suspension of ethanol liquid containing the particles. Dispersion of droplets containing single particles diffuse in the vacuum chamber. Once a particle is trapped the vacuum chamber is evacuated by a pumping station (HiPace 80 Pfeiffer, 70 l/s). In our procedure, the pressure is first decreased to $10^{-3}$~mbar, with an optical trapping power of at least $400$~mW. The vacuum chamber returns to atmospheric pressure and back to low pressure several times. If the scattering intensity has not changed, it means that particle may have been densified and pure fused silica is obtained. This procedure allows us to keep the particle trapped indefinitely from $1$~atm to $10^{-3}$~mbar.

\section{Nonlinear Calibration}

\begin{figure}[h]
\begin{center}
\includegraphics[width=135mm]{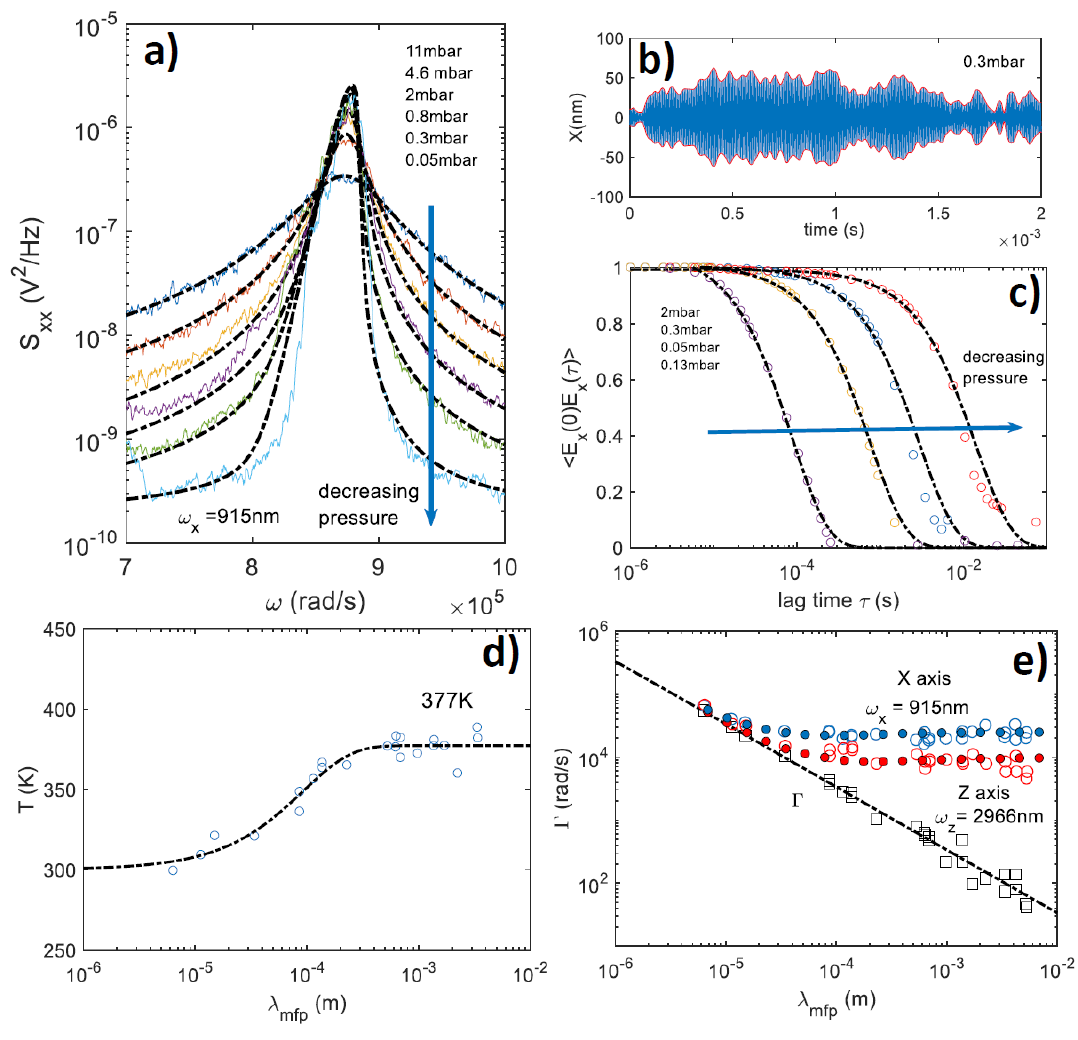}
\end{center}
\caption{a) Experimental power spectral densities $S_{xx}$ along the transverse $x$
axis for several pressures computed from the time traces of the particle's motion. The dashed lines correspond to fits to the data
using the theoretical expression Eq.~(\ref{PSDR}). b) Typical experimental time trace along the $x$
axis. c) Normalized energy correlation functions with exponential fits. d)
Center of mass motion temperature $T$\ versus mean free path $\lambda
_{\text{mfp}}$. e) Effective damping rates measured either by looking at the
FHWM of the PSD along $x$ (blue circles) and $z$ axis (red circles) or the
inverse of the correlation time in c) (black squares). The dashed line is a
fit to the data using Eq. (\ref{G}) with a single fitting parameter
$R_{p}=68nm$. The blue and red filled circles correspond to the FHWM of the
PSD\ of the theoretical model [see Eq. (\ref{PSDR})].}
\label{FigS2}
\end{figure}

We compute PSDs from time traces of the particle's motion along the three axes (with a
typical noise floor $\sim$pm$^{2}$/Hz). \ For clarity, we only present results
for the $x$ transverse and $z$ axial components. Fig. \ref{FigS2}(a) displays PSDs for
several pressures. Upon lowering the pressure, the system crosses from the
quasi-linear to the nonlinear regime [22]. We observe a
clear transition towards nonsymmetric lineshapes below roughly 1 mbar induced
by nonlinearities.

It is necessary to, first, quantify the three-dimensional stochastic Duffing
oscillator both for calibration purposes but also to infer the beam parameters
$w_{j}$'s used for analyzing the role of the radiation pressure and to
determine the center of mass motion temperature. For weak displacements
$x_{i}=x,$ $y$ or $z$ of the particle, relative to the trapping laser
wavelength $\lambda$, we calculate to first order in the $x_{i}$ the optical
forces assuming the Gaussian Beam approximation. A
softening of the effective stiffness $\kappa_{i}^{\text{eff}}(\mathbf{x}%
)=\kappa_{i}\left(  1-\sum_{j=x,y,z}\xi_{ij}x_{j}^{2}\right)  $, \emph{i.e}
due to nonlinearities, in the gradient force $F_{i}^{\text{grad}}=-\kappa
_{i}^{\text{eff}}(\mathbf{x})x_{i}$ is experimentally observed through the
frequency shift of the linear resonance $\Omega_{i}=\left(  \kappa
_{i}/m\right)  ^{1/2}$, with $m$ the mass of the particle. This leads to asymmetric profiles in the PSDs rather than
the usual Lorentzian arising in the linear regime. Here
$\kappa_{i}= \alpha' I_0/\left( w_i^2 (1+\delta_{iz}) \right)$ is the linear trap stiffness, seen by the coordinate $x_{i}$, while $I_0$ represents the field intensity at the focus. Duffing nonlinearities are quantified by the
coefficients $\xi_{ij}=2/w_{j}^{2}$ except for $\xi_{zj=x,y}=4/w_{j}^{2}$. The theory of a 1D underdamped anharmonic oscillator undergoing random
thermal motion has been already described in [23,24]. To
calculate spectral densities [24], we adapt this theory to the 3D case
with the secular approach. The theoretical
analysis is based on two parameters, the damping rate $\Gamma$ and an
anharmonicity parameter $\alpha_{i}^{R}$ quantifying the strength of
nonlinearities $\alpha_{i}^{R}=\frac{3}{4}C_{i}\xi_{ii}k_{\mathrm{B}}%
T/m\Omega_{i}\Gamma$ with $C_{i=x,y}=9/4$, $C_{i=z}=7/2$ and
where $k_{\mathrm{B}}$ is the Boltzmann constant. This analysis is valid in
the underdamped ($\Gamma\ll\Omega_{i}$) and low-temperature regimes
($\alpha_{i}^{R}\Gamma\ll\Omega_{i}$).

The Duffing oscillator's analysis uses the damping coefficient $\Gamma$ of the particle in a rarefied gas taken to be [S1]
\begin{equation}
\Gamma=\frac{6\pi\eta R_{p}}{m}\frac{0.619}{0.619+Kn}(1+c_{K}),\label{G}%
\end{equation}
where $R_{p}$ is the particle's radius and $\eta$ the viscosity of air. The
dependence of $\Gamma$ with the gas pressure $P$ depends on the Knudsen number
$Kn=\lambda_{\text{mfp}}/R_{p}$, where $\lambda_{\text{mfp}}$ $=\lambda
_{\text{mfp0}}\cdot P_{\text{atm}}/P$ is the mean free path of air molecules
with $P_{\text{atm}}$ and $\lambda_{\text{mfp0}}=68$ nm being respectively the
atmospheric pressure at $20{{}^{\circ}}C$ and the corresponding mean free
path. The additional correction in $Kn$ is given by $c_{K}%
=(0.31Kn)/(0.785+1.152Kn+Kn^{2})$. In the limit $Kn\gg1$, we find
$\Gamma\propto P$. 

In terms of the hypergeometric function $_{2}F_{1}$, the
spectral density of the fluctuations of the coordinate $x_{i}$ is then given
by [24]%
\begin{equation}
S_{ii}(\omega)=\frac{k_{\mathrm{B}}T}{m\Omega_{i}^{2}}\sum_{l=\pm
1}\operatorname{Re}\left\{
\left(  1+q_{i}\right)  ^{-4}\frac{16q_{i}}{\Gamma\left(  \kappa_{l}^{i}(\omega)+1\right)}{}
_{2}F_{1}\left(  2,\kappa_{l}^{i}(\omega)+1,\kappa_{l}^{i}(\omega)+2;\xi
_{i}
\right)\right\},\label{PSDR} 
\end{equation}
with $q_{i}=(1+4i\alpha_{i}^{R})^{1/2}$, $\kappa_{l}^{i}(\omega)=-\left(
\Gamma/2-i\left[  l\omega+\Omega_{i}-\Gamma^{2}/8\Omega_{i}\right]  \right)
/q_{i}\Gamma$, and $\xi_{i}=-\left(  1/4\alpha_{i}^{R}\right)  ^{2}\left(
1-q_{i}\right)  ^{4}$. We calibrate the optical tweezer at relatively high
pressures ($T=300K)$, that is, in the weakly nonlinear regime $\left(
\alpha_{i}^{R}\ll1\right)  $ where the spectral density is given by%

\begin{equation}
S_{ii}(\omega)=\frac{2k_{\mathrm{B}}T}{m}\frac{\Gamma_{i}^{e}}{\left(  \left[
\Omega_{i}^{e}\right]  ^{2}-\omega^{2}\right)  ^{2}+\left(  \omega\Gamma
_{i}^{e}\right)  ^{2}}.\label{PSDL}%
\end{equation}
Eq. (\ref{PSDL}) corresponds to a single Lorentzian line with renormalized
resonance frequency $\Omega_{i}^{e}$ $=\Omega_{i}+2\alpha_{i}^{R}\Gamma
-\Gamma^{2}/8\Omega_{i}$ and the linewidth $\Gamma_{i}^{e}$ $=\Gamma\left(
1+4\left[  \alpha_{i}^{R}\right]  ^{2}\right)  $.\ 

The calibration is performed as follows:

\textbf{1)} We fit the PSDs at high pressure with the Lorentzian form given in
Eq. (\ref{PSDL}), where they are indeed well approximated by a Lorentzian
form, to determine the effective eigenfrequencies $\Omega_{i}^{e}$ and damping
$\Gamma_{i}^{e}$. Using the nominal nanoparticle size given by the
manufacturer to estimate $\Gamma$ via Eq.~(\ref{G}) then provides a first
estimate for $\alpha_{i}^{R}$.

\textbf{2)} The extraction of both the local envelope amplitude $\langle
x^{2}(t)\rangle^{1/2}$ and the local frequency $\Omega_{x}(t)$ allows for a
measurement of the time resolved energy potential associated with the
direction $x$, $E_{x}(t)=m\Omega_{x}^{2}(t)x^{2}(t)/2$. Its correlation
function decreases exponentially as expected with a characteristic time
$\Gamma^{-1}$ as shown in Fig.~\ref{FigS2}(c). The size of the
particle is then verified independently using Eq.~(\ref{G}). We find a
particle radius $R_{p}=68$ nm (\textit{i.e.} the nominal size given by the manufacturer).

\textbf{3)} Fitting\textbf{\ }Lorentzian PSDs with effective eigenfrequencies
$\Omega_{i}^{e}$ and damping $\Gamma_{i}^{e}$ at high pressure allows the
calibration from volts $V$ to meters for each axis. This is via the
determination of the prefactor $2k_{B}T/m$ in Eq.~(\ref{PSDL}) where (i) we
take $T=300K$ in the high pressure data as the gas density is assumed
sufficient to equilibrate the particle temperature with that of the
surrounding gas (ii) $m$ is determined from $R_{p}$ and the density of
$SiO_{2}$ given earlier. We then use Eqs.~(\ref{PSDR}) to fit the spectral
densities [Fig.~\ref{FigS2}(a)] between the regimes of weak and strong nonlinearities,
where the slight discrepancy seen in the wings of PSDs are due to a lack of
statistics at low pressure and to the $10^{-3}$ relative laser power
fluctuations. This step is performed iteratively to minimize the error between
the experimental Full Half Width Maximum (FHWM) of PSDs and FHWM given by the
nonlinear theory Eqs.~(\ref{PSDR}).

\textbf{4) }The temperature $T$ of the system in the low pressure regime is
deduced relative to the ambient temperature  (\emph{i.e.} 300 K as reference
at high pressure for $\alpha_{i}^{R}<1$), for which the calibration has been
carried out using Eqs.~(\ref{PSDR}-\ref{PSDL}) as shown in Fig.~\ref{FigS2}(d).

The final trap parameters obtained from this procedure are:~$w_{x}=0.915$%
~$\mu$m, $w_{y}=1.034$~$\mu$m and $w_{z}=2.966$ $\mu$m. The reason why $w_x\neq w_y$ is due to the high numerical aperture of the objective. Wave vectors with high angles generate new polarisations and the field distribution at the focus is slightly elongated along the direction of the polarisation of the incident field [S2]. Fig.~\ref{FigS2}(e) clearly
shows a saturation of the linewidths $\Gamma_{i}^{e}$, that is well reproduced
by the full height maximum widths FHMW given by Eq.~(\ref{PSDR}). As soon as
$\alpha_{i}^{R}>1$, the widths deviate from each other such that $\Gamma
_{x}^{e}>\Gamma_{z}^{e}$, meaning that $\alpha_{x}^{R}>\alpha_{z}^{R}$. The
knowledge of these geometrical parameters of the focused laser beam are
crucial for the analysis of the influence of scattering forces, as demonstrated in the main text.

\end{document}